\documentclass[fleqn,usenatbib]{mnras}

\usepackage{mathptmx}

\usepackage[T1]{fontenc}
\usepackage{ulem}
\usepackage{color}    
\DeclareRobustCommand{\VAN}[3]{#2}
\let\VANthebibliography\thebibliography
\def\thebibliography{\DeclareRobustCommand{\VAN}[3]{##3}\VANthebibliography}

\usepackage{graphicx}	
\usepackage{amsmath}	
\usepackage{amssymb}	
\usepackage[switch]{lineno}

\title[Neutron star LMXBs with long recurrence times]{A large population of neutron star low-mass X-ray binaries with long outburst recurrence time ?}

\author[E. Meyer-Hofmeister, H. Cheng  and B.F. Liu]{E. Meyer-Hofmeister$^{1}$, \thanks{emm@MPA-Garching.MPG.DE} 
Huaqing Cheng,$^{2}$
and B.F. Liu $^{2,3}$\\
 $^{1}$ Max-Planck-Institut f\"ur Astrophysik, Karl-Schwarzschild-Str.~1, D-85741 Garching, Germany\\
$^{2}$ Key Laboratory of Space Astronomy and Technology, National Astronomical Observatories, Chinese Academy of Sciences, Beijing 100101, China\\
$^{3}$ School of Astronomy and Space Science, University of Chinese Academy of Sciences, 19A Yuquan Road, Beijing 100049, China
}
\date{Accepted XXX. Received YYY; in original form ZZZ}

\pubyear{2023}
\begin{document}
\label{firstpage}
\pagerange{\pageref{firstpage}--\pageref{lastpage}}
\maketitle

\begin{abstract}
Low-mass X-ray binaries (LMXBs) with neutron stars 
    show quite different features which depend on the rate of mass transfer 
    from the donor star.
    With a high transfer rate the Z sources are in a persistent soft
    spectral state, with a moderate rate the transient Atoll sources have
    outburst cycles like the black hole X-ray binaries. The
    observations document very long outburst recurrence times for quite
    a number of sources.
    We follow with our computations the evolution of the accretion disc
    until the onset of the ionization instability. For sources with a
    low mass transfer rate the accumulation of matter in the disc is
    essentially reduced due to the continuous evaporation of matter
    from the disc to the coronal flow. Different
    mass transfer rates result in nearly the same amount of matter
    accumulated for the outburst which means the outburst properties
    are similar for sources with short and sources with long outburst cycles,
    contrary to some expectations. Then of systems with long recurrence
    time less sources will be detected and the total population of
    LMXBs could be larger than it appears. This would  relieve the
    apparent problem that the observed number of LMXBs as progenitors
    of  millisecond pulsars (MSP) is too small compared to the number
    of MSP.  Concerning the few quasi-persistent sources with year-long
    soft states we argue that these states are not outbursts, but
    quasi-stationary hot states as in Z sources.
 \end{abstract} 


\section{Introduction}

In low-mass X-ray binaries  matter is transferred from a low-mass star
onto a black hole or a neutron star. The mass
overflow can be caused by the evolution of the donor
star or loss of angular momentum due to either gravitational radiation or 
magnetic braking in binaries with short orbital periods. In most cases
matter flows via an accretion disc toward the compact object. 

The vertical structure of an accretion disc  in close
binary stars is determined by the amount of matter accumulated in
the outer disc and can be either
a hot or a cool state. For a certain range both states are
possible and the ionization instability can trigger the transition
from an initially cool state into a hot state, the outburst. This
instability was first understood for accretion onto a white dwarf in
cataclysmic variables \citep{mmh1981},  
and later recognized as essential for the evolution of X-ray binary
stars \citep{mhr1993}.
In low-mass X-ray binaries (LMXB) the matter is accreted onto a black hole or
a neutron star. The mass transfer rates depend on the mass and the
evolutionary state of the secondary star. A detailed description of the computation of the vertical structure,
the disk evolution and the instability behaviour of 
black hole and neutron star LMXBs is given in
the review of \citet{lasota2001} for a wide range of parameters.
A recent review of the evolution of  neutron star
LMXBs leading to the
formation of millisecond pulsars is given by \citet{dantona2022}.

 An important process for disc evolution, especially in discs
 around a black hole or a neutron star, is the evaporation of matter
 from the disc to the corona  \citep{mmh1994,lbf1999,mlmh2000a,mlmh2000b}. If the rate of mass transfer from the
 secondary star is low the evaporation of matter from the inner disc
 can lead to an extended truncation of the inner disc appearing as a
 hard spectral state. 
 This leads to outburst recurrence times of decades of years.

There are differences between  black hole LMXBs and neutron star
 LMXBs. A possible magnetic field in neutron star LMXBs, if strong
 enough, can  cause a truncation of the inner disc within the Alf\'ven
 radius \citep[e.g.][]{frank2002} and influence the outburst
 behaviour. The irradiation from the surface of the neutron star is
 essential and its effect was early discussed by
 \citet{vanpar1996,king1996,king1997} and \citet{burderi1998};
 Different luminosities are observed in LMXBs.  For black hole LMXBs
the observations document large outbursts  
with luminosities of $10^{37}-10^{39} \rm{erg \,s^{-1}}$
after long quiescence with low luminosities of
$10^{30} - 10^{33} \rm{erg \,s^{-1}}$ \citep{yanyu2015, tet2016}.
 In neutron star binaries the
outburst peak luminosities are about a factor of 10 lower than those
of black hole sources and quiescent levels are typically near
$10^{32}-10^{33}\, \rm{erg \,s^{-1}}$ \citep{tom2004}.

 The neutron star LMXBs show different phenomena
  according to the mass transfer from the secondary star. With the
highest accretion rates Z sources are persistently in a soft
spectral state. With also high rates the bright atoll sources remain
almost permanently in the bright soft state \citep{mundar2014}. 
Transient atoll sources have lower accretion rates and are observed to
change between soft and hard state in outburst cycles. 
\citet{mundar2014} studied in their investigation
of hysteresis the spectral states of 50
neutron star binaries monitored by the Rossi X-ray Timing
Explorer (RXTE), including also sources with very long recurrence time
or long lasting outbursts. Recently \citet{macc2022} discussed the
observation of transient neutron star LMXBs and pointed out that about
half of the known sources have recurrence times in excess of a decade
(for outbursts at the sensitivity of MAXI) and argued that a much
larger number of sources would not be discovered. With his investigation he
clarified that long recurrence times are not only observed for
black hole LMXBs. This was suggested by \citet{hai2018}, but already in
contradiction to an example of a binary evolution computed by
\citet{dub2001}. This gives rise to a
study of the accretion disc evolution in neutron star LMXBs to
understand the long recurrence times and the outbursts of these sources.

The aim of our computations is to study the disc evolution in binaries
with long recurrence times. We describe in Sect.2 the computational
procedure which is used to determine the accumulation
of matter in the disc. In Sect.3 we
show the results for disc evolution and 
 reveal that during the quiescent
phases evaporation of matter from the
disc to the coronal flow  
plays an important role.
The consequences of evaporation
for the recurrence times and outburst strength and duration are discussed in
Sect. 4. In Sect.5 we compare these results with observations for
binaries with long recurrence times.
In contradiction to perhaps expected powerful
outbursts after a long lasting accumulation of matter the observed
outburst luminosities are comparable to those in systems with short
outbursts cycles. This affects the discovery rate of LMXBs and is of
importance for the total number of LMXBs. In the context of the
recyling scenario \citep{dantona2022} the population of LMXBs as the
progenitors of MSPs should be comparable with the population of MSPs. 
But too few LMXBs are observed. In Sect.6 we try to explain the
nature of very long outbursts of quasi-persistent X-ray transients
\citep{wij2003}, which are probably not outburst cycles, but sources in a hot
quasi-stationary state of the disc. Conclusions follow in Sect.7.

\section{Modelling the accretion disc evolution} 

The structure of accretion discs in neutron star LMXBs
shows similar features as that of discs around black holes and also in
cataclysmic variables around white dwarfs. A standard modelling is
adequate including the gravity of the compact primary star and 
additional features as the effect of a neutron star magnetic field and
the radiation from the neutron star surface. Taking into account the
influence of irradiation on the transient nature the ionization
instability is expected as the cause for the outburst behaviour for
all these binaries \citep{vanpar1996}.

In quiescence the
structure of accretion discs around a neutron star or a black hole
primary for most binaries consists of a geometrically thin disc
together with a hot corona, an inner ADAF
(advection-dominated accretion flow, developed by Narayan, for a
review see \citet{nar1998}). As studied originally for discs in
cataclysmic variables and applied to black
hole LMXBs matter can be evaporated  from the cool thin disc to a
coronal flow in a ``siphon-flow'' process \citep{mmh1994}. The
evaporation then leads to the truncation of the thin disc, 
which depends on the mass flow rate in the thin
disc. If the accretion rate is low the innermost part is completely
evaporated into an ADAF and the spectrum
is hard. Otherwise,  the disc reaches inward to the last stable orbit
and the spectrum is soft.

Our main computations concern the accumulation of matter in
quiescence. We take the computer code as composed for studying the
long recurrence times of outbursts in black hole LMXBs, especially the
disc evolution of A0620-00 \citep{mhm1999}. We use
for the cool state a viscosity parameter 0.05, which led for A0620-00 to
a good agreement of the total matter accumulated and the recurrence time with
observations (for a discussion of very low
values see Sect.5). For the viscosity in the hot state we use the value 0.2.
The physics of the interaction of disc and corona, which causes the
evaporation of matter from the disc to the corona, was included as
described in \citet{lbf1997}. The critical values of surface density
$\Sigma_A(r)$ and $\Sigma_B(r)$ for an unstable accretion disc structure
are taken from the relations derived by \citet{lud1994}. These relations were
used for the computation of the evolution of the disc in cataclysmic
variables \citep{lbf1997} and black hole binaries \citep{mhm1999}.
\begin{equation}
    \begin{array}{l}
    \Sigma_A(r) = 32\ \rm{g\, cm^{-2}}\, \left(\frac{M_{1}}{M_\odot}\right)^{-0.37}
  \left(\frac{r}{10^{10}cm}\right)^{1.10} \left(\frac{\alpha_{\rm  h}}{0.2}\right)^{-0.74},\\
    \Sigma_B(r) =185\rm{g\,cm^{-2}}\, \left(\frac{M_{1}} {M_\odot} \right)^{-0.37} \left(\frac{r} {10^{10}cm}\right)^{1.10}\left(\frac{\alpha_{\rm c}}{0.05}\right)^{-0.8}
    \end{array}
\end{equation}
with  $M_{1}$ the primary star mass, $r$ the disc radius,
  $\alpha_{\rm h}$ and $\alpha_{\rm c}$ the viscosity values
 for the hot and the cool
state respectively.

During the disc evolution the inner radius is determined by the
evaporation process within short time, independent of the initially
assumed value.
The outer boundary of the disc is given by the
tidal radius or for small mass ratios by the 3:1 resonance
between Kepler binary period and Kepler rotation in the disc, which ever
is smaller \citep{frank2002, menn2016}. The
resonance transfers any surplus of angular momentum
outflow in the disc to the orbit \citep{white1988, lub1991}.
The extension of the disc 
depends on the mass ratio. For the neutron star mass we take 1.4$M_{\odot}$.
We assumed the masses of the secondary stars 
based on the investigation of orbital period changes during evolution by
\citet{pibi2002} (for short periods only auxiliary).  

The accumulation of matter in the disc during a long lasting
quiescence  essentially  depends on the evaporation process:
the corona is fed by matter from the disc underneath, so
that an equilibrium establishes between the cool accretion stream and
the hot flow. We use the results derived by \citet{liuf1995} applied
to LMXBs \citep{mhm1999}.
\begin{equation}
  \dot M_{\rm{evap}}
  =10^{-10.2}M_{\odot}/\rm{yr}\ \left(\frac{M_1}{M_\odot}\right)^{2.34} \left(\frac{r}{10^9{\rm cm}}\right)^{-1.17}\  
\end{equation}

With evaporation the evolution of the surface density is governed by conservation of mass and angular momentum,
\begin{equation}\label{e:diffusion}
    \frac{\partial{\Sigma}}{\partial{t} }=\frac{3} {r} \frac{\partial}{\partial{r}} \left[{r^{1/2} \frac{\partial} {\partial{r}}\left( r^{1/2}\nu \Sigma  \right)}\right]-\dot m_{\rm evap},
\end{equation}
where $\nu$ is the kinematic viscosity, $\dot m_{\rm evap}$ is the evaporation rate per unit surface area, which is related to the integrated evaporation rate $\dot M_{\rm evap}$  \citep[see][]{lbf1997}.
When the surface density reaches $\Sigma_B$ an outburst is triggered.

Additional effects in disc evolution are caused by irradiation or a magnetic field of the neutron star.

 The effect of irradiation from the neutron
star surface on the outburst behaviour is difficult to include as
shown in the detailed review by \citet{dub2001}. It was shown that
irradiation is negligible for the accumulation of matter in quiescence
with low accretion rates. But irradiation has an effect on the
outburst behavior and  on the recurrence time because after an
irradiation-controlled outburst the amount of matter in the disc is
low \citep{dub2001}. We assume for our computations that only a small
amount of matter is left over in the disc after the outburst. We take the
initial distribution $\Sigma(r) =0.1 \,\Sigma_A(r)$. 
For low mass transfer rates the surface density in the
outermost region then can be too low for the propagation of
the heating front to the outer edge of the disc, and due to the
matter remaining in the cool outermost part the quiescent 
time until the next outburst would be somewhat shorter.

 An additional feature of neutron star LMXBs is
a magnetic field of the neutron star, which also can cause a
truncation of the disc within the Alf\'ven radius
\citep[e.g.][]{frank2002}. Such a disc truncation was discussed for 4U
1608-52 by \citet{veijn2020}. Moreover it is possible that the
accreting matter is forced to follow the magnetic field lines of a
rapidly rotating magnetic neutron star known as propeller effect
\citep{ill1975}. The efficiency of the propeller effect was
discussed by \citet[][and references therein]{men1999} 
for neutron star and black hole sources.
It depends on the magnetic field strength and the spin
velocity of the neutron star and further assumptions of the
properties of the binary. The fraction of matter finally accreted on the
surface of the neutron star could be so low that this effect is more
important than evaporation. For black hole sources without a solid
surface only the
possibility remains that the wind existing with the coronal flow
reduces the matter accretion onto the compact object. A further process
limiting the accretion of matter onto the neutron star could be,
together with a stop of the mass transfer from the secondary star,
the ignition of the radio pulsar which sweeps away the mass at the
inner Lagrangian point as discussed by \citet{burderi2001}.

\section{Computation of the recurrence time}

The accumulation of matter in the disc depends on the rate of mass transfer
from the secondary star and only low or very low mass transfer rates
will lead to long quiescent phases (see \citet{dub2001} for a LMXB
example). Low mass transfer rates are confirmed theoretically as
shown by \citet{ritter1999} who gave an analytical formula for the rate
of nuclear driven mass transfer from a Roche lobe filling red
giant. Observations document very low accretion rates in some cases as listed by
\citet{watts2008} in their investigation of LMXBs on the selection for
the Advanced LIGO observing run.

For the computation of the evolution of the accretion disc during quiescence
we have chosen the parameters 
 neutron star mass and disc size derived for  MXB 1659-298, a source for
which a very low X-ray flux was clearly detected by \citet{wij2003}. 
MXB 1659-298 is a source with a long recurrence time of 14 years:
  After the launch of RXTE \citep{lin2019} two
outbursts 1999/2001 \citep{wij2003} and 2015/2017 \citep{iar2019}
 occurred. The orbital period
is 7.11 hours, typical within the range of neutron star LMXBs as
described by \citet{lin2019}. 
For our computations we take 1.4 $ M_{\odot}$ for the mass of
the neutron star, 0.7 $ M_{\odot}$ for that of the secondary star, and an orbital
period of 7 hours, with which the size of disc is determined. The neutron
star mass could be higher as pointed out by \citet{oez2016} and by \citet{watts2008}. The
secondary star mass of $0.7M_\odot$ is derived from the study of \citet{pibi2002},
which was also considered by \citet{watts2008} to lie in the range between 0.1 and
0.78 $M_{\odot}$ and by \citet{iar2018} in the range between 0.3 and 1.2
$M_{\odot}$. 

With the chosen parameters for the neutron star and secondary star mass, the orbital period, and  the mass transfer rate at the outer boundary, the diffusion equation (\ref{e:diffusion}) is integrated with an assumed initial radial distribution of surface density \citep[for details see][]{lbf1997}, and thus the evolution of the matter
distribution in the disc can be determined as shown in the following
figures.

Fig.\ref{f:accum1} shows how the matter distribution in the disc
evolves until the surface density reaches the critical value
${\Sigma_B}$ for the onset of the instability. For a low accretion rate
the evaporation of matter to the coronal flow
causes a quite large inner hole filled with the ADAF. With the accretion
rate of $\dot M_{\rm{transfer}}=5.52\, 10^{-11} M_{\odot}/\rm{yr}$ the outburst
is triggered after a recurrence time of 14 years at the distance of
$r=\rm{10^{9.7}}$cm. 
In the inner region, $r\la 10^{9.4}$cm,  only
a corona/ADAF flows toward the neutron star  as a consequence of complete evaporation. 

\begin{figure}
\centering
\includegraphics[width=8.0cm]{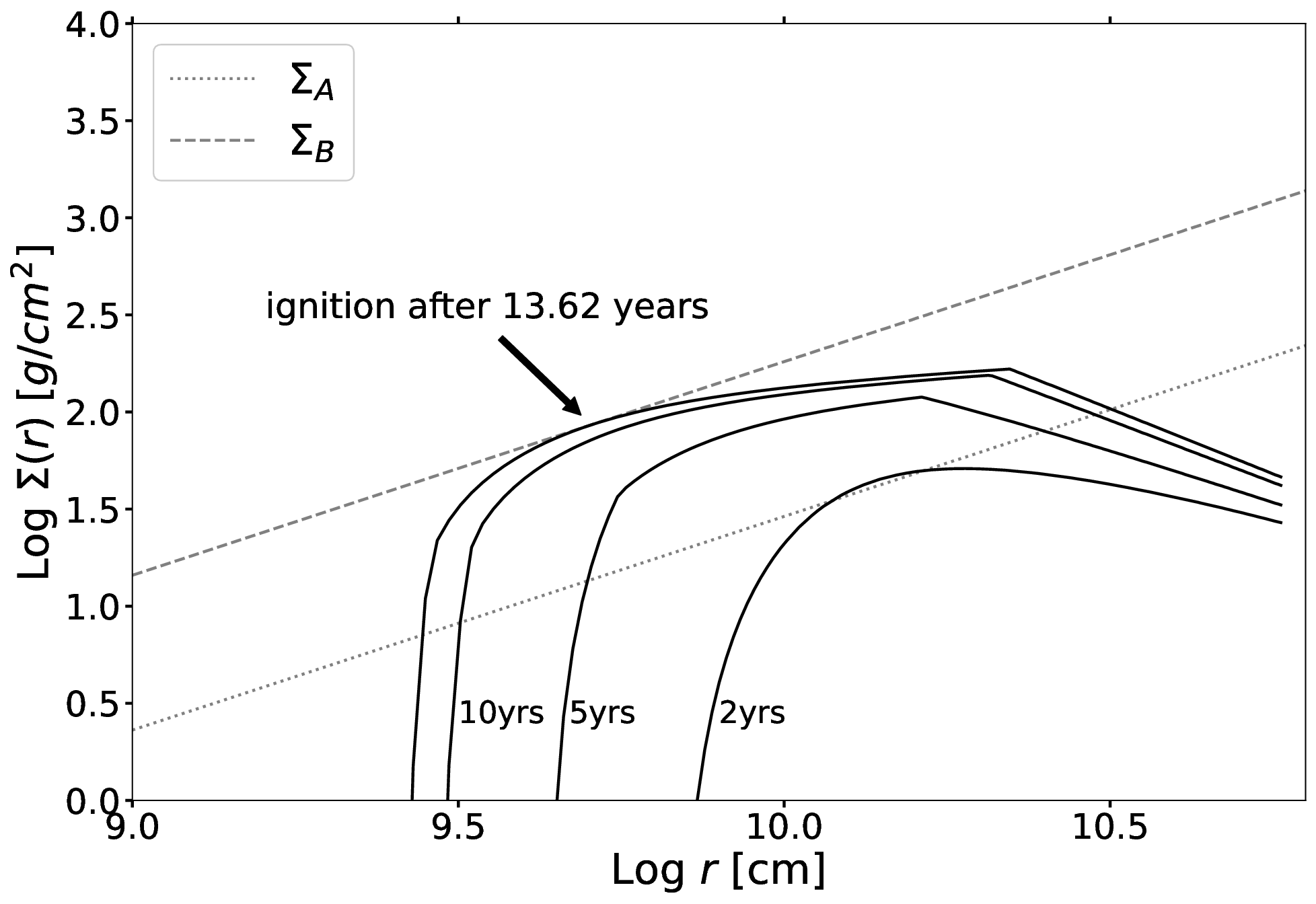}
      \caption{Modelling MXB 1659-298, evolution of the disc surface
       density during the quiescent state,
       $\dot M_{\rm{transfer}}=5.52\,10^{-11} M_{\odot}/\rm{yr}$,
       $\rm{M_1}$=1.4 $M_{\odot}$,\, $\rm{M_2}$=0.7$M_{\odot}$,
       ${\rm{P_{orbital}}=\rm{7hrs}}$)
       }
    \label{f:accum1}
\end{figure}

\begin{figure}
\centering
\includegraphics[width=8.0cm]{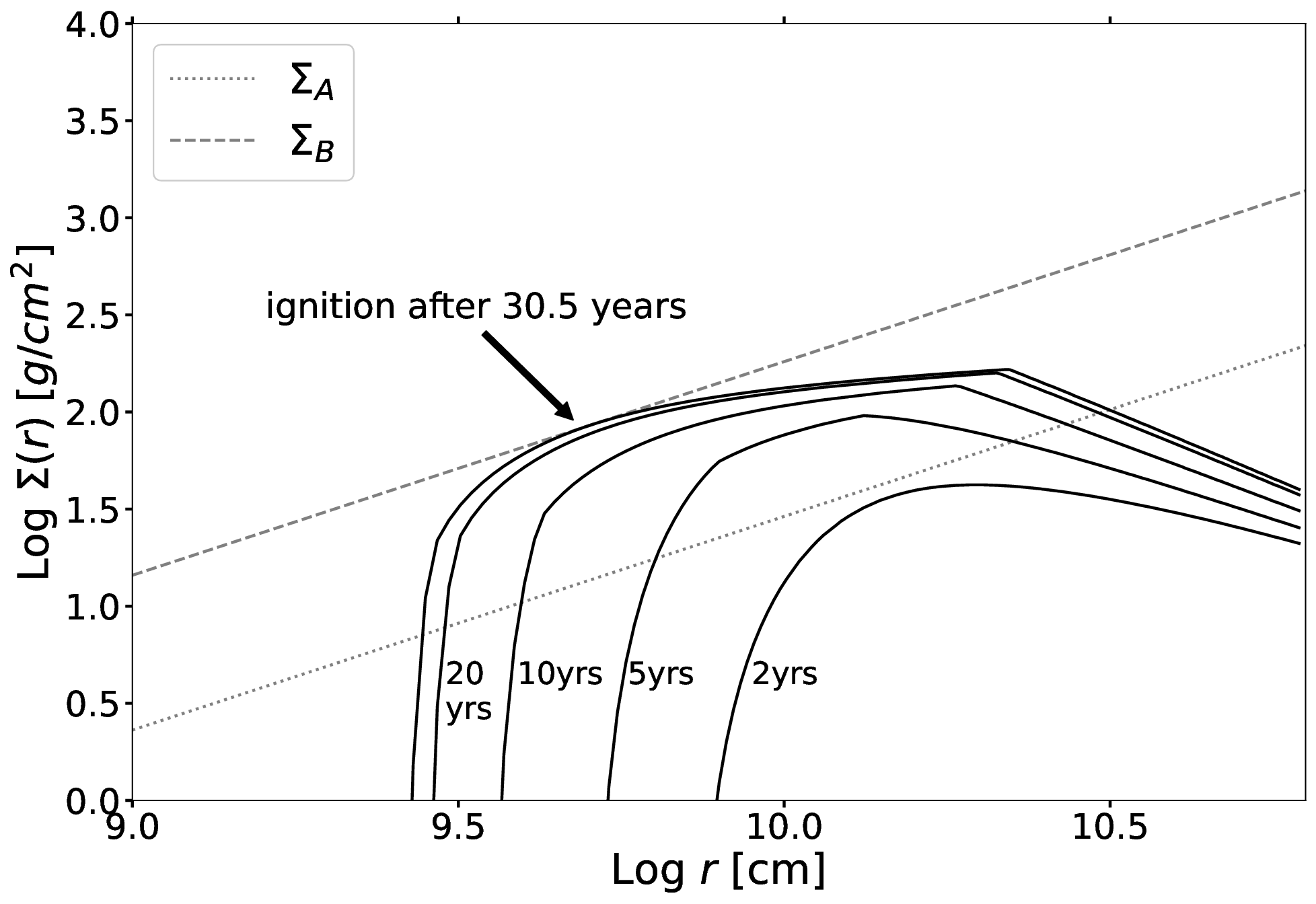}
    \caption{Evolution of disc surface density during the quiescent
      state, $\dot M_{\rm{transfer}}=4.6\,10^{-11}
      M_{\odot}/\rm{yr}$, other parameters as in Fig.\ref{f:accum1}.}
     \label{f:accum2}
\end{figure}

Fig.\ref{f:accum2} shows the accumulation of matter for the lower
accretion rate of $\dot M_{\rm{transfer}}=4.6\,10^{-11} M_{\odot}/\rm{yr}$. The
balance between storage of matter in the disc and
evaporation establishes an inner hole of about the same extension as
for the higher rate. The critical surface density to trigger an outburst
is reached at about the same distance as with the higher 
rate. The recurrence time is longer, about 31 years. 
The time needed to reach the instability is very long since the
chosen mass transfer rate is only slightly higher than the
evaporation rate at the truncation radius.

\begin{figure}
\centering
\includegraphics[width=8cm]{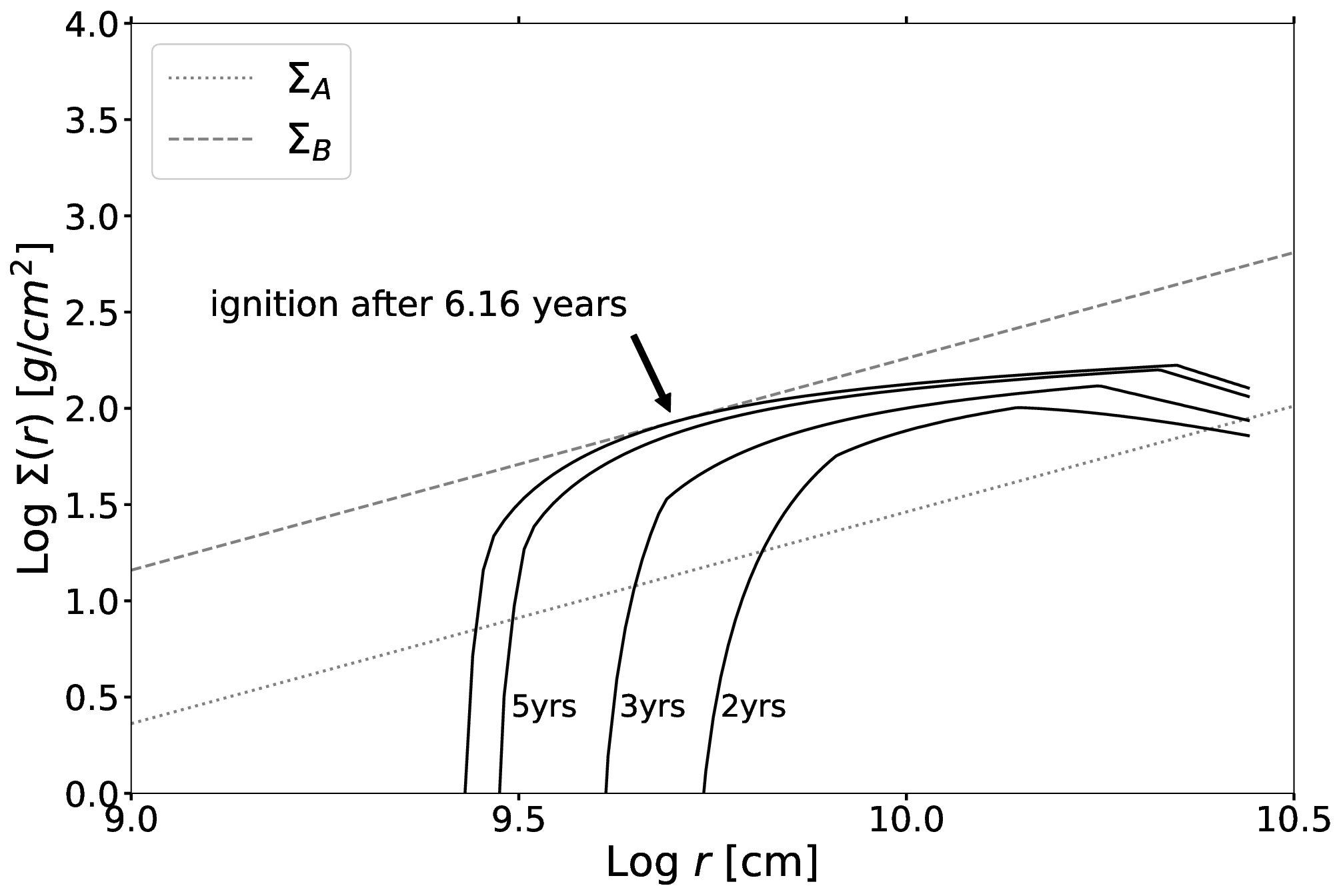}
\caption{
  Evolution of disc surface density during the quiescent
      state, $\dot M_{\rm{transfer}}=5.52\,10^{-11} M_{\odot}/\rm{yr}$, $\rm{M_1}$=1.4
      $M_{\odot}$,\, $\rm{M_2}$=0.3$M_{\odot}$,
       ${\rm{P_{orbital}}=\rm{3hrs}}$
       }
       \label{f:accum3}
\end{figure}

In neutron star LMXBs with shorter orbital periods the accumulation of matter can lead earlier to an outburst.
To study this effect we compute the disc evolution also for an orbital
period of 3 hours and and we take again 
 the mass transfer rate of $\dot M_{\rm{transfer}}=5.52\, 10^{-11}
 M_{\odot}/\rm{yr}$ (used for the accumulation shown in Fig.\ref{f:accum1}).  We
 take a smaller secondary star mass,
0.3$M_{\odot}$.  The smaller mass ratio leads to a limitation of the
disc size due to the 3:1 resonance (see Sect.3). The recurrence time
then is much shorter, only
6.16 years. The accumulation of matter for this case is shown in
Fig.\ref{f:accum3}. Generally for binaries with shorter orbital
periods, where the two components are closer and the
discs are smaller, enough  matter to trigger an outburst can be
accumulated within shorter time.

For a binary with a longer orbital period, e.g. 12 hours,
the recurrence times are longer since more matter is stored in the
larger disc until an outburst is triggered. 
But independent of the disc size evaporation reduces the accumulation of
matter.
 
\begin{figure}
\centering
\includegraphics [width=8.0cm]{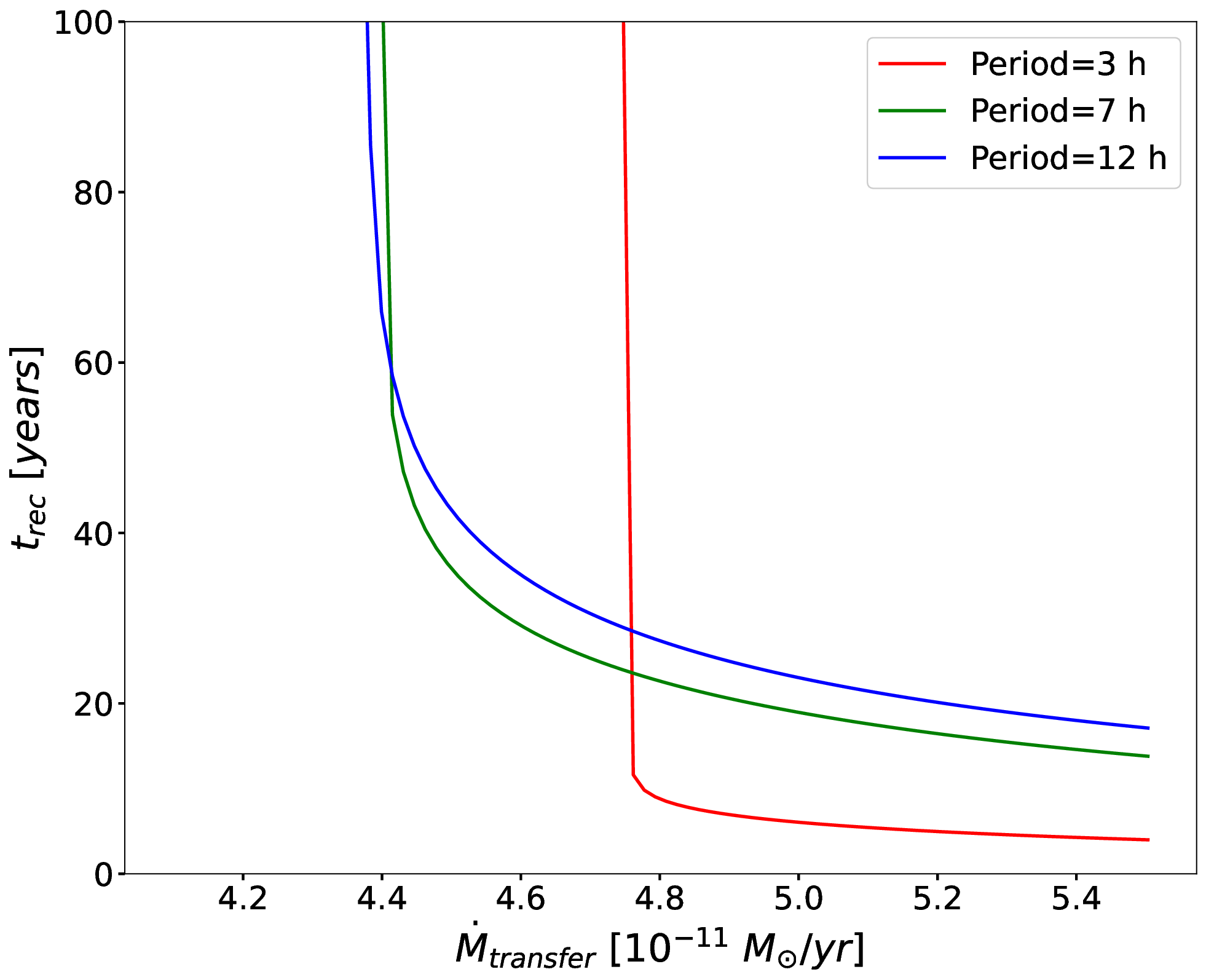}
    \caption{Dependence of the recurrence time on mass transfer rate
      and orbital period}
    \label{f:rec}
\end{figure}

Fig.\ref{f:rec} shows how the recurrence times for different orbital
periods depend
on the mass transfer rate. A recurrence time of a few years would be
expected for quite a range of mass transfer rates, but for low rates
a much longer time since only a small fraction of the transferred mass is accumulated in the disc. For even lower rates the transferred mass is balanced by the evaporation and thus the source could stay in the
stationary cool state without an outburst. The vertical lines in
Fig.\ref{f:rec} give the limiting rate for the occurrence of the
ionization instability.

The onset of an outburst from a stationary cool state might in
rare cases be triggered in a
different way: For sources with mass transfer rates so low that they
remain always in the cool state, random fluctuations of mass transfer
could cause a transition to the hot state. 
Such a change of the disc structure then
appears as an outburst after long recurrence time.

We have not included magnetic fields of neutron stars for our
modelling of the accretion disc structure. Using observations \citet{asai2016}
investigated the effect of neutron star magnetic fields of LMXBs and
found that in general the magnetic fields are weak. X-ray spectral
observations, especially new data from NuSTAR (Nuclear Spectroscopic
Telescope Array) and NICER (Neutron star Interior Composition Explorer)
for several sources, e.g. for atoll sources \citep{ludl2019}, led to
the determination of the inner edge of the accretion disc and the
determination of the magnetic field strength responsible for this
truncation. The computation of the disc evolution for the parameters
corresponding to a considered binary then would allow to investigate
whether evaporation or the effect of the magnetic field are more
important.  \citet{deg2017} pointed out that the disc truncation
determined by modelling the reflection spectrum of J17062-6143 could be
due to either evaporation of the inner disc into an ADAF or the
pressure of the neutron star magnetic field which could exhibit a
strong radio jet as well as a (propeller-driven) wind-like outflow .

\section{Distinct outburst properties predicted by the evaporation model}
          
The evaporation is an important process that 
influences the accumulation of matter during quiescence and the
outburst behaviors. The most distinct property is that
during quiescence matter continually evaporates from the disc, 
leading to a longer accumulation time until an outburst is triggered.   This effect is
particularly important at low transfer rates as it takes away a
significant fraction of the transferred gas flow during the evolution.
If the rate is so low that all the transferred gas can be
evaporated,  no outburst can be triggered. On the other hand short
recurrence times are expected for high transfer rates where the
evaporation only takes away small fraction of the transferred matter
and the influence on the accumulation is small.

The effect of evaporation on the total amount of matter stored in the
disc when the instability sets in is essential for the outburst
behavior of LMXBs with low mass transfer rates.
Our computations result in the following amount of accumulated matter:
For the higher mass transfer rate of $5.52\,10^{-11}
M_{\odot}/\rm{yr}$ an amount of $9.16\,10^{23}$g,
for the lower rate of $4.6\,10^{-11} M_{\odot}/\rm{yr}$ an amount of 
$9.52\,10^{23}$g. It is interesting that about the same amount of
matter is accumulated in the disc for the different mass transfer
rates (see the distributions of the surface density at ignition in
Figs. \ref{f:accum1} and \ref{f:accum2}). Otherwise one finds
the total amount of matter transferred from the secondary
star to the disc, $1.5\,10^{24}$g for the higher rate and about double
the amount for the lower rate during the longer quiescence. This means
that a fraction of 61\% of the transferred matter is accumulated in
the disc for the higher rate and only 34\% for the lower rate.
The  evaporated gas flows inwards as an ADAF, of
  which a significant
  fraction is lost in wind at large distances \citep{yuanf2014} and
  the remaining part could flow along magnetic lines at small
  distances or accrete directly to the neutron star depending on the
  strength of magnetic field and the spin.
Eventually a propeller
  effect can determine the flow of the matter.
  
The accumulation of matter in black hole LMXBs is reduced
by evaporation in the same way, during also long quiescent phases
as computations for A0620-00 show \citep{mhm2000}, with a similar
fraction of finally accumulated matter in the disc though for
different parameters. Due to a higher mass transfer rate 
and the larger size of the disc then a bright outburst occurs
\citep{lloyd1977}. 

The outcome of the same amount of matter accumulated 
in the discs of different neutron star LMXBs at the onset of
the instability in spite of  different (low) transfer rates implies  
about the same outburst strength, peak luminosity and duration,
contrary to what might be expected due to the long lasting accumulation of
matter. This means that the systems with long recurrence
times are visible in outburst only during a shorter time and one
might expect that many LMXBs with long recurrence time are not yet detected.

We want to point out that the effect of the viscosity parameter
is distinct from that of evaporation.  A smaller viscous parameter
means slower accretion and thus more gas accumulation in the outer
region until an outburst can start. With a low transfer rate the
accumulation takes also a long time, but the large amount of gas accumulated in the outer disc could lead to outburst properties different from those
caused by evaporation, though both of which can predict a long
recurrence time.

In the context of viscosity one might ask whether the
long recurrence times in LMXBs could be caused in a similar way as
those in WZ Sagittae stars, a subgroup of cataclysmic variables, where
the primary star is a white dwarf. But the large amount of matter
accumulated in these binaries demands a very low viscosity. A
sharp enhancement of mass transfer is excluded (for a discussion see
\citet{smak1993, mhm1998, mmh1999, ham2000}).

\section{Comparison with observations}

We want to compare with observations for long recurrence times and with
the properties of the outbursts after a long recurrence time.

\citet{macc2022} selected a sample of 20 transients
and came to the conclusion that at least 40\% of transient LMXBs
have recurrence times longer than 10 years. \citet{lin2019} investigated the
outburst rate based on the X-ray monitoring data from Swift/BAT,
RXTE/ASM and MAXI in the  past few decades and found that 10 of
the 19 neutron star LMXBs had only one outburst. \citet{watts2008}
summarized the properties of 35 accreting neutron stars in our Galaxy
for the detection of gravitational wave emission and derived estimates
for the average long-term and outburst flux. They found for most
of the sources with rare outbursts an outburst duration not
particularly long, only 10 to 60 days (as also observed for other
transient sources \citep{mundar2014}) and peak luminosities not
especially high, only around 1/10 of the peak luminosity of the
well-known source Aql X-1.
  The average outburst flux is 1.3\,$10^{-8}   
\rm{erg \, cm^{-2}s^{-1}}$ \citep[Table
1]{watts2008} which corresponds to a luminosity of    
3.2 - 3.7 $10^{37} \rm{erg \,s^{-1}}$ for the usually assumed   
distance of 4.5-5 kpc (the outburst history is shown 
in the work of \citet{
Ootes2018}).

A subgroup of LMXBs with, for most sources,
only one observed outburst since their discovery during the last 20
years are the accretion powered millisecond pulsars (AMXP). The
general properties were reviewed by \citet{patruno2021} and \citet{disalvo2022}.The outburst luminosities were faint and most sources did
not not reach a soft spectral state. The outbursts of AMXPs last
from less than a week up to about 60 days in most sources \citep{mar2019}.
These observations are in good agreement with the theoretically
derived expectations for the outburst behaviour for
the amount of accumulated matter in the disc during quiescence. Only
the source HETE J1900.1-2455, also an AMXP, seems to have a remarkable
complex activity different from the other sources.

It is possible that the mass transfer is non-conservative
in LMXBs. \citet{mar2019} investigated such a situation for accreting
millisecond pulsars and found strong evidence for this in five sources out of
his sample of 18 sources.

\section{The disc structure in quasi-persistent transients}

Besides outburst cycles with long recurrence time a few neutron star
LMXBs show a definitely different long-term behaviour. These sources,  designated as
quasi-persistent sources \citep{wij2003}, are observed in a soft spectral state
for years with an about constant luminosity. Detailed
observations document the long outbursts of the three sources EXO
0748-676, KS 1731-260 and HETE J1900.1-2455 lasting 23, 12.5 and 10
years (the outburst duration might even be longer since the sources were
discovered already in outburst). With the final transition to a lower
luminosity the sources then enter a continuous quiescent state. The
luminosity decrease between these two states is less than the decay in
regular outburst cycles of neutron star LMXBs. It was for EXO 0748-676
a factor of 4 \citep[Fig.1]{mundar2014}. For the understanding of the accretion process also the quiescent state is of interest, for HETE J1900.1-2455 discussed by \citet{degen2021}  and for EXO 0748-676 by \citet{pari2021}.

What is the nature of these sources?  The computation of matter
accumulation in the disc shows that during quiescence due to
evaporation only a limited amount can be stored, insufficient 
for a very long outburst.  Therefore continually enough matter must be transferred
from the secondary star to the disc at a high rate,
i.e. that the sources are in a quasi-stationary hot state. 
This is the case also in other sources within the regime of
neutron star LMXBs, known as bright atolls or Z sources, which remain
almost permanently in a bright soft state. The source
XTE J1701-462 even displayed a change from Z and atoll phenomenology
to a regular outburst \citep{homan2007,homan2010}.

A steady soft state is possible if the mass transfer
rate from the secondary star is high enough so that  the entire disc
is at the upper, stable branch of the S-shaped curve of the 
viscosity-surface density relation
\citep{lud1994}. A limit cycle
behavior appears when the mass transfer rate lies within some
range.  At this range the local unstable behavior of some annuli
influences its neighbors and produces a coherent behavior of the
entire disc,  leading to outburst cycles with no need of variation
of the transfer rate.   
A  change from a soft spectral state to a hard
state  can also be triggered if the mass transfer rate  decreases from  sufficiently high to low values. 
When the surface density decreases with the decreasing transfer rate  to a value below the
critical surface density $\Sigma_A(r)$,  usually first fulfilled at the outer disc edge, a cooling
front moves inward and establishes a cold disc structure.
This probably happens in quasi-persistent sources and is caused by a
temporary decrease of the mass transfer rate after a long lasting hot
state.

In the long lasting hot state the surface density $\Sigma(r)$
increases inward and the disc extends down to the ISCO.
The surface density distribution during the
quasi-stationary hot, soft state is different from that in the hard 
state (in quiescence) shown in Figs.1 to 3, 
where the inner region is filled by an ADAF. A minimal mass transfer
rate  $\dot M_A(r)$ which allows a stationary hot state can be
determined as in the work of \citet[Fig.13]{lud1994}
corresponding to the relations for the critical values of surface
density. The limiting rate depends on the disc size (therefore
on the mass of the secondary star).

For a binary with a short orbital period of 3 hrs and a secondary star mass 
of 0.3$M_{\odot}$ (example in Sect. 3) the disc size is
limited by the 3:1 resonance and the limiting mass transfer rate is lower.
Then for low transfer rates more likely a quasi-stationary state can exist.

This might be the
case for the source with the shortest orbital period HETE J1900.1-2455
\citep{simon2018} and an average value of the mass transfer rate during
the 10 years of outburst of $3.5\ 10^{-10} M_\odot$/yr \citep{deg2017}. It was
claimed by \citet{pap2013} that the source behaves as a typical atoll source. 
The properties of EXO 0748-676 are similar, with the longest observed
soft state of 23 years, an orbital period of 3.82 hrs and an accretion rate of
0.022 $\rm{\dot M_{Edd}}$ over the course of the RXTE observations,
which is about $7\ 10^{-10} M_\odot$/yr \citep{gall2008}. For both of these sources
the rates determined from the observations are probably high enough to allow a stationary hot state.

\section{Conclusions}
  
We investigate the evolution of the accretion disc until the onset of
an outburst for binaries with long recurrence time and therefore low mass
transfer rates from the secondary star. The existence of low rates is clearly
confirmed by observations \citep{watts2008,jon2006}. A significant
result of our computations of disc evolution for low
mass transfer rates is the essentially decreased accumulation
of matter in the disc due to evaporation of matter from
the geometrically thin, optically thick disc to a coronal flow toward
the neutron star. 

It is important that different mass transfer rates
lead to nearly the same amount of finally accumulated matter for the outburst.
This means that the outburst after the long quiescence has only a standard peak
luminosity and duration in most sources similar or less than of the outbursts in sources with
shorter recurrence time, e.g. Aql X-1 \citep{camp2013}. Due to the
occurrence of only standard outbursts after long time, maybe after 10 or 20
years, less such sources will be detected and quite a number remain
undetected and it is difficult to determine the extent of the
population of neutron star LMXBs.

Since in the context of the recycling scenario \citep{disalvo2022}
neutron stars are spun up to millisecond pulsars (MSP) the number of observed
LMXBs as progenitors of MSPs (taken the birthrate of LMXBs) should be
comparable with this population. But the population is smaller at
least a factor of 2.5 \citep{dantona2022} or even a factor of 5
\citep{macc2022}. This ``birthrate problem'' would be dimineshed if a
large number of undetected LMXBs exist as already claimed by
\citet{macc2022} and as our computations suggest.

The main differences between the neutron star LMXBs and the black hole LMXBs are whether there exist a solid surface and magnetic fields. In addition to the relevant effects discussed in Sect.2,  there is a new effect concerning the evaporation. During quiescence the evaporated gas flows inwards as an ADAF with an accretion rate $\sim 10^{-11}M_\odot/{\rm yr}$. This hot flow is mostly lost in wind at large distances as shown by MHD simulations \citep[for a review see Sect.3.4 of ][]{yuanf2014}, with the accretion rate decreasing  with decreasing distance in a power law.  Further diversion of the accreting flow occurs in the inner region, affected by the magnetic field. Therefore, the mass flows to the neutron star surface could be at a rate less than  $\sim 10^{-13}M_\odot/{\rm yr}$, which results in a luminosity from the boundary layer of the solid surface within the observational range. Uncertainty in this estimation of the quiescent luminosity lies in the true fraction of evaporated gas flowing to the neutron star surface. 

Concerning the year-long outbursts of quasi-persistent X-ray
transients we want to point out that the so-called outbursts are a
feature very different from long recurrence times, so to speak an
opposite state. In quasi-persistent sources
the mass transfer from the secondary star seems to be strong enough to
maintain a soft spectral state for a long time, that is to keep the
  surface density above the critical value $\Sigma_A$ in the curve of
the viscosity-surface density relation everywhere in the
disc. As shown such a quasi-stationary hot state is more likely possible
for sources with a short orbital period where the disc is smaller.
The source HETE J1900.1-2455 with the shortest orbital period of 1.39 hrs and
EXO 0748-676 with an orbital period of 3.82 hrs might be
examples. Further observations of long-term phenomena might help
to better understand these features.

\section*{acknowledgements}

 We thank the referee for detailed comments and
suggestions to improve the paper.
Financial support for this work is provided by the National Natural
Science Foundation of China (Grant No. 12073037,  No. 12203071, and No.12333004).

\section*{Data availability}
The data and code used in this work will be shared on reasonable
request to the corresponding authors.


 \bibliographystyle{mnras}
 \bibliography{ns} 



\bsp	
\label{lastpage}
\end{document}